\documentclass{ws-ijmpb}
\usepackage{graphicx}%

\begin{document}
\markboth{Swarnali Bandopadhyay, A. M. Jayannavar}
{Phase time for a tunneling particle}

\def\f{\frac}
\def\rdown{\rho_{\downarrow}}
\def\pa{\partial}
\def\th{\theta}
\def\Ga{\Gamma}
\def\ka{\kappa}
\def\bea{\begin{eqnarray}}
\def\eea{\end{eqnarray}}
\def\be{\begin{equation}}
\def\ee{\end{equation}}
\def\pa{\partial}
\def\d{\delta}
\def\K{\kappa}
\def\a{\alpha}
\def\eps{\epsilon}
\def\th{\theta}
\def\na{\nabla}
\def\nn{\nonumber}
\def\lan{\langle}
\def\ran{\rangle}
\def\pr{\prime}
\def\rarrow{\rightarrow} 
\def\larrow{\leftarrow}

\title{Phase time for a tunneling particle}
\vskip 20 mm
\author{Swarnali Bandopadhyay}
\address{Department of Physics, Ben-Gurion University, Beer-Sheva 84105, Israel\\
swarnali@bgu.ac.il} 
\author{A. M. Jayannavar}
\address{Institute of Physics, Sachivalaya Marg, 
Bhubaneswar 751005, India\\
jayan@iopb.res.in}

\date{\today}

\maketitle

\begin{abstract} 
We study the nature of tunneling phase time for various quantum 
mechanical structures such as networks and  
rings having potential barriers in their arms. We find the generic 
presence of Hartman effect, with superluminal velocities as a 
consequence, in these systems.
In quantum networks it is possible to control the `super arrival'
time in one of the arms by changing the parameters on another arm which is 
spatially separated from it. This is yet another quantum nonlocal effect. 
Negative time delays (time advancement) and `ultra Hartman effect' with 
negative saturation times have been observed in some parameter regimes.
In presence and absence of Aharonov-Bohm (AB) flux quantum rings show
Hartman effect. We obtain the analytical expression for the
saturated phase time. In the opaque barrier regime this is 
independent of even the AB flux thereby generalizing the Hartman effect. 
We also briefly discuss the concept of ``space collapse or 
space destroyer" by introducing a free space in between two barriers 
covering the ring. Further we show in presence of absorption the 
reflection phase time exhibits Hartman effect in contrast to the 
transmission phase time. 
\end{abstract}
\vskip 0.5cm 
\keywords{ Tunneling,  Electronic transport}
PACS numbers: 03.65.-w; 73.40.Gk; 84.40.Az; 03.65.Nk; 73.23.-b 


\section{Introduction}
\label{time1}
Quantum tunneling, where a particle has finite probability to penetrate a
classically forbidden region is an important feature of wave mechanics.
Invention of the tunnel diode\cite{tsu73}, the scanning tunneling microscope\cite{bining86} etc. made the quantum tunneling useful from a technological 
point of view. In 1932
MacColl\cite{mac32} pointed out that tunneling is not only characterized
by a tunneling probability but also by a time the tunneling particle
takes to traverse the barrier. There has been considerable interest on
the question of time spent by a particle in a given region of space\cite{landauer94,hauge89,anantha02}. The recent developments of nanotechnology 
brought new urgency to study the tunneling time as it is directly related
to the maximum attainable speed of nanoscale electronic devices. In addition, 
recent experimental results claiming superluminal tunneling speeds for
photons\cite{steinberg,nimtz} call for detailed analysis of this problem. In a number of numerical\cite{num}, experimental\cite{steinberg,expt} and analytical study of quantum 
tunneling processes, various definitions of tunneling times have been 
investigated. These different time scales are based on various different 
operational definitions and physical interpretations.
Till date there is no clear consensus about the existence of a simple 
expression for this time as there is no hermitian operator associated with it\cite{landauer94}. Furthermore, a corpuscular picture of tunneling is very
hard to be realized due to the lack of a direct classical limit for the
 trajectories and velocities of the tunneling particle.

Among the various time scales, `{\it dwell time}'\cite{smith59} which gives 
the duration of a particle's stay in the barrier region regardless of how it 
escapes can be calculated as the total probability of the particle inside the
barrier divided by the incident probability current. The 
`{\it conditional sojourn time}'\cite{anantha02} gives the time of 
sojourn (dwell) in the spatial region of interest for some given conditions 
of scattering. It can be defined 
meaningfully by a `clock' which is basically an extra degree of freedom 
that co-evolves with the sojourning particle. 
B\"uttiker and Landauer proposed\cite{buttiker82} that one should study 
`tunneling time' using the transmission coefficient through a static 
barrier of interest, supplemented by a small oscillatory perturbation. 
This is referred to as `{\it B\"uttiker-Landauer time}' in the literature.
In another approach, the
traversal time, is measured by the spin precession of the tunneling particle 
in a uniform infinitesimal magnetic field. This is called local `{\it Larmor 
time}'\cite{smith59,buttiker83prb}. A large 
number of researchers interpret the `{\it phase time}'\cite{hauge89,wigner55,hartman62} as the temporal delay of a transmitted wave 
packet. 
This time is usually taken as the difference between the time at which the peak of the transmitted packet leaves the barrier and the time at which the peak of the incident quasi-monochromatic wave packet arrives at the barrier. 
Within the stationary phase approximation, 
the phase time can be calculated 
from the energy derivative of the `phase shift' in the transmitted or 
reflected  amplitudes. It would be worth mentioning that for a symmetric
barrier, the reflection and transmission phase times are equal. For tunneling 
through a $1$D static barrier due to the time-reversal symmetry, one can show from the unitarity of the scattering matrix that the phases of the reflection and transmission amplitudes differ by a constant, quantitatively by $\pi/2$. 
However this does not remain true for asymmetric/complex barrier. 
B\"uttiker-Landauer\cite{buttiker82} raised objection 
that the peak is not a reliable characteristic of packets as the wave packet may undergo strong distortion or deformation after tunneling through the barrier. 
Moreover, there is no causal relationship between the peak of the transmitted 
wave packet and the peak of the incident packet. This is due to the fact that
the peak of the transmitted packet can leave the scattering region before the peak of the incident packet has arrived.
In contrast to `dwell time' which can be defined locally, the
`phase time' is essentially asymptotic in character\cite{hauge87}. 
The `phase time' statistics is intimately connected with dynamic
admittance of micro-structures\cite{buttiker93}. The `phase time' is
also directly related to the density of states\cite{buttiker02,avishai,swarnali02,swarnali03}. 
The universality of `phase time' distributions in random and chaotic
systems has already been established earlier\cite{jayan}. In the case of `not
too opaque' barriers, the tunneling time evaluated either as a simple 
`phase time'\cite{hauge89} or calculated through the 
analysis of the wave packet behaviour\cite{recami92} becomes independent of the
barrier width. This phenomenon is termed as the Hartman effect 
\cite{hartman62,recami92,fletcher}. This implies that for sufficiently long 
barriers the effective velocity of the particle can become arbitrarily large, 
even larger than the light speed in the vacuum (superluminal effect). Though
this interpretation is a little far fetched for non-relativistic Schr\"odinger 
equation as velocity of light plays no role in it, this effect
has been established even in relativistic quantum mechanics.

Tunneling time associated with electron being very small ($\sim $ femto second) the experiments to verify the `Hartman effect' (generalized Hartman effect)
are usually done on optical waveguide where the corresponding time is of the
order of pico second. However
the formal analogy between the Schr\"odinger equation and 
the Helmholtz equation for electromagnetic wave enables one to 
correlate the results for optical experiments 
to that for electrons. Photonic experiments show 
that electromagnetic pulses travel with group velocities in excess of the 
speed of light in vacuum as they tunnel through a constriction in a
waveguide\cite{nimtz}. Experiments with photonic band-gap structures 
clearly demonstrate that `tunneling photons' indeed 
travel with superluminal group 
velocities\cite{steinberg}. Their measured tunneling time is practically 
obtained by comparing the two peaks of the incident and transmitted wave 
packets. Thus all these experiments directly or indirectly
confirmed the occurrence of Hartman effect without violating  
`Einstein causality' {\em i.e.}, the signal velocity or the information 
transfer velocity is always bounded by the velocity of light. 
It should  also be noted that in the photonic tunneling time experiments
by Nimtz {\em et. al.}, based on frustrated total internal reflection, the velocity
of the half-width of the pulse (not the peak of the wave packet) is monitored.
The velocity of the half-width is found to be superluminal (according to
theory and experiment). Relation of this result to the causality principle
is discussed in the references\cite{nimtz99,nimtz01}. 
The `Hartman effect' has been extensively 
studied both for nonrelativistic (Schr\"odinger equation) and relativistic 
(Dirac equation)\cite{landauer94,hauge89,nimtz} cases. Recently Winful 
\cite{winful03} showed that the saturation of phase time is a direct 
consequence of saturation of integrated probability density under the barrier 
(equivalently in the electromagnetic waves saturation of stored energy).
The Hartman effect has been found in 
one dimensional barrier tunneling\cite{recami92} 
as well as in tunneling through mesoscopic rings 
in presence of Aharonov-Bohm (AB) flux\cite{swarnali04}. In the present work 
we extend the study of phase times for branched quantum networks 
and rings.

The main results of this paper are as follows:
\begin{itemize}
\item We have found Hartman effect in barrier tunneling regime for quantum networks and rings.
\item In quantum networks non-locality and time-advancement (negative time delay) are found.
\item In presence of AB-flux in the rings saturated phase time becomes independent of the flux. Thus the Hartman effect is obtained even in presence of AB-flux.
\item We obtain space collapse in quantum ring i.e. in presence of inter-barrier free space the phase time becomes independent of the length of this free space in the off resonance regimes.
\item Even in presence of absorption in the barrier, the reflection phase time shows Hartman effect though the transmission phase time grows with the length of the barrier.
\end{itemize}

The paper is organized as follows. In section~\ref{time3} we present
 phase times\cite{swarnali05} for branched networks of quantum wires 
which can readily be 
realized in optical wave propagation experiments. This geometry allows us
to check other nonlocality effect such as tuning of the saturation value of
`phase time' and consequently the superluminal speed in one branch by changing
barrier strength or length in any other branch, spatially separated from the 
former. In section~\ref{time2} we study the Hartman effect on a quantum ring 
geometry 
{\it i.e.} beyond one dimension and in the presence of 
AB-flux\cite{swarnali04,swarnali04proc}. 
Our results confirm `Hartman effect' in quantum ring even in presence of 
AB-flux. 
Then we study the phase time in quantum ring in absence of AB-flux. In tunneling through the ring having two potential barriers with an intermediate free space where a quantum particle can propagate as a traveling wave, the saturated phase time becomes independent of the intermediate free length (in the large length 
limit of the barriers) in the off resonant cases. This result can be 
interpreted as a ``space collapse or space destroyer"\cite{recami02}, as if
the intermediate free space does not exist. We also show that in the presence
of absorption in barrier regime the Hartman effect survives in the reflection 
mode as opposed to the transmission mode. Then in section~\ref{conclude} we
 summarize our results and conclude.
 
\section{Hartman effect and non-locality in quantum networks}\label{time3}
\begin{figure}[t]
\begin{center}
\includegraphics[width=10.0cm]{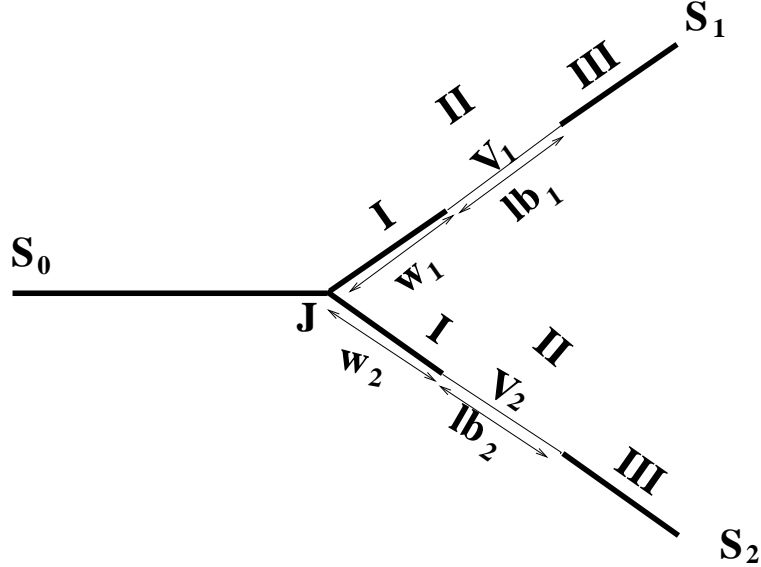}
\end{center}
\caption{Schematic diagram of a Y-junction or three-way splitter.} 
\label{system}
\end{figure}

Hartman effect is itself one of the manifestations of quantum non-locality
\cite{steinberg}. Here we study the effect for various quantum mechanical 
networks having potential barriers in their arms. In such systems 
it is possible to control the `super arrival' time in one of the arms by 
changing  parameters on another, spatially separated from it. This is yet 
another quantum nonlocal effect. Negative time delays (time advancement) 
and `ultra Hartman effect' with negative saturation times have been 
observed in some parameter regimes.

As a model system we choose a network of 
wires. The length of these
wires are so narrow that only the motion along the length of the wires
is of interest (a single channel case). The motion in the perpendicular 
direction is frozen in the
lowest transverse sub-band. In a three-port Y-branch circuit 
(Fig.~\ref{system}) the two side branches of quantum wire $S_1$ and $S_2$ are 
connected to a `base' arm $S_0$ at the
junction $J$. In general one can have $N (\geq 2) $ such
side branches connected to the `base' wire.

We study the scattering problem across a network geometry as
 presented schematically in Fig.~\ref{system}. 
Such geometries are important from the point of view of basic science due to  
their properties of tunneling and interference\cite{washburn86,gefen84} 
as well as of technological applications 
such as wiring in nano-structures. In particular, the Y-junction carbon 
nanotubes are in 
extensive studies and they show various interesting properties like 
asymmetric current voltage characteristics\cite{cnr}. 
In our system of interest there are finite 
quantum mechanical potential barriers of strength 
$V_1$ and $V_2$ and length $lb_1$ and $lb_2$ in the side branches 
$S_1$ and $S_2$ respectively. 
We focus on a situation wherein 
the incident electrons have an energy $E$ less than $V_n$, $n=1,2$. 
The impinging electrons in this sub-barrier 
regime travel as an evanescent mode/wave and the transmission involve 
contributions from quantum tunneling and multiple
reflections between each pair of barriers and the junction point. 
Here we are interested in a single channel case where the 
Fermi energy lie in the lowest sub-band. To excite the evanescent modes 
in the side branches one has to produce constrictions by making the 
length of the regions of wires containing barriers much thinner than 
that of the other parts of the wire. The electrons 
occupying the lowest sub-band in the connecting wire on entering the 
constrictions 
experience a potential barrier (due to higher quantum zero point energy)
and propagate as an evanescent mode\cite{jayan94mpl,gupta}.
In this work an analysis of the phase time or 
the group delay time in such a system is carried out.

\subsection{Theoretical treatment}  
We approach this scattering problem using the quantum wave guide theory 
\cite{xia92,deo94}. In the stationary case the incoming particles
are represented by a plane wave $e^{ikx}$ of unit amplitude. The
effective mass of the propagating particle is $m$ and the energy is 
$E=\hbar^2 k^2/2m$ where $k$ is the wave vector corresponding to the free 
particle. The wave functions,
in different regions of the system considered in  Fig.~\ref{system} can be 
written as,
\begin{eqnarray}
\psi_{in}(x_0) &=& e^{ikx_0} + R e^{-ikx_0}\, \, \, \,
 (\mbox{in}\, S_0) \, ,\nn\\
\psi_{(1)_{I}}(x_1) &=& A_1\, e^{i\,k\,x_1} + B_1\, e^{-i\,k\,x_1}\,\,\,\,(\mbox{region I in}\, S_1) \, , \nn\\
\psi_{(2)_{I}}(x_2) &=& A_2\, e^{i\,k\,x_2} + B_2\, e^{-i\,k\,x_2}\,\,\,\,(\mbox{region I in}\, S_2) \, , \nn\\
\psi_{(1)_{II}}(x_1) &=& C_1\, e^{-\kappa_1 \,(x_1-w_1)} + D_1\, e^{\kappa_1 \,(x_1-w_1)}\,\,\,\,(\mbox{region II in}\, S_1) \,,\nn\\
\psi_{(2)_{II}}(x_2) &=& C_2\, e^{-\kappa_2 \,(x_2-w_2)} + D_2\, e^{\kappa_2 \,(x_2-w_2)}\,\,\,\,(\mbox{region II in}\, S_2) \,,\nn\\
\psi_{(1)_{III}}(x_1)&=& t_1\, e^{i\,k\,(x_1-w_1-lb_1)}\,\,\,\,(\mbox{region III in}\, S_1) \,,\nn\\
\psi_{(2)_{III}}(x_2)&=& t_2\, e^{i\,k\,(x_2-w_2-lb_2)}\,\,\,\,(\mbox{region III in}\, S_2) \,,\nn
\end{eqnarray}
with $\kappa_n=\sqrt{2m(V_n-E)/\hbar^2}$ being the   
imaginary wave vector in presence of a rectangular barrier of
strength $V_n$, with $n=1,2$. $x_0$ is the spatial coordinate for the 
`base' wire, whereas $x_1$ and $x_2$ are the spatial coordinates for 
the $S_1$ and $S_2$ side branches respectively.  
 All these coordinates are measured from the junction $J$. $w_1$ and $w_2$ 
are the distances between the junction $J$ and the respective barriers in
arm $S_1$ and $S_2$.

To solve the problem, we use Griffith's boundary conditions\cite{griffith}
\begin{equation}
\psi_{in}(x_0=J)=\psi_{(1)_{I}}(x_1=J)=\psi_{(2)_{I}}(x_1=J),\label{bc1}
\end{equation}
and 
\begin{equation}
\frac{\partial \psi_{in}(x_0)}{\partial x_0}\Big |_J = 
\frac{\partial \psi_{(1)_{I}}}{\partial x_1}\Big |_J \, + 
\frac{\partial \psi_{(2)_{I}}}{\partial x_2}\Big |_J \,,
\label{bc2}
\end{equation}
at the junction $J$. All the derivatives are taken either outward or inward 
from the junction\cite{xia92}. In each side branch, at the starting and end 
points of the barrier, the boundary conditions
can be written as
\begin{eqnarray}
\psi_{(n)_{I}}(x_n=w_n)=\psi_{(n)_{II}}(x_n=w_n)\, ,\label{bc3} \\
\psi_{(n)_{II}}(x_n=w_n+lb_n)=\psi_{(n)_{III}}(x_n=w_n+lb_n)\, ,\label{bc4} \\
\frac{\pa \psi_{(n)_{I}}}{\pa x_n}\Big |_{(w_n)} = 
\frac{\pa \psi_{(n)_{II}}}{\pa x_n}\Big |_{(w_n)} \, ,\label{bc5} \\
\frac{\pa \psi_{(n)_{II}}}{\pa x_n}\Big |_{(w_n+lb_n)} = 
\frac{\pa \psi_{(n)_{III}}}{\pa x_n}\Big |_{(w_n+lb_n)} \, ,\label{bc5} 
\end{eqnarray} 
where $n=1,2$.
From the above mentioned boundary conditions one can obtain 
the complex transmission amplitudes $t_1$ and $t_2$ on the side branches
$S_1$ and $S_2$ respectively. 
\subsection{Results and Discussions}

Following the
method introduced by Wigner\cite{wigner55}, we can calculate the 
`phase time' (phase time for transmission) from the energy derivative of 
the phase of the transmission amplitude $t_n$\cite{hauge89,wigner55} as
\begin{equation}
\tau_n = \hbar \, \frac{\partial Arg[t_n]}{\partial E}\, . 
\label{phtm}
\end{equation}

In what follows, we set $\hbar =1$ and $2m =1$. We now proceed to analyze
the behavior of $\tau_n$ as a function of various 
physical parameters for different network topologies. We measure
time at the far end of each barrier in the branched
arms containing barriers and in the case of arms in absence of any barrier
we measure the phase time at the junction points. We express all the
physical quantities in dimensionless units {\em i.e.} all the barrier strengths 
$V_n$ in units of incident energy $E$ ($V_n \equiv V_n/E$), all the
barrier lengths $lb_n$ in units of inverse wave vector 
$k^{-1}$ ($lb_n \equiv k lb_n$), where 
$k=\sqrt E$ and all the extrapolated phase time $\tau_n$ in units of inverse 
of incident energy $E$ ($\tau_n \equiv E\tau_n$).

\begin{figure}[t]
\begin{center}
\includegraphics[width=10.0cm]{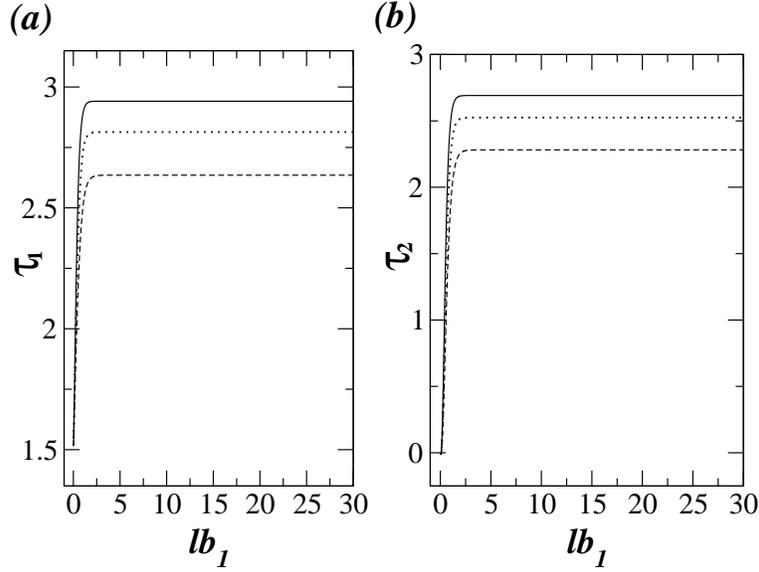}
\caption{For a 3-way splitter with a barrier in $S_1$ arm, the `phase times' $\tau_1$ and $\tau_2$ are plotted as a function of barrier length `$lb_1$' in $(a)$ and $(b)$ respectively. The solid, dotted, dashed curves are for $V_1=5, 4, 3$ respectively. Other system parameters are $ E = 1, w_1 = 3$.} 
\label{1barrier}
\end{center}
\end{figure}

First we take up a system similar to the Y-junction shown in 
Fig.~\ref{system} in presence of a 
barrier $V_1$ of length $lb_1$ in arm $S_1$ but in absence of any barrier
in arm $S_2$. For a tunneling particle having energy $E < V_1$ we find
out the phase time $\tau_1$ in arm $S_1$ as well as $\tau_2$ in arm $S_2$ 
as a function of barrier length $lb_1$ (Fig.~\ref{1barrier}).
From Fig.~\ref{1barrier}(a) it is clear that $\tau_1$ evolves with $lb_1$ and 
eventually saturates to $\tau_{1s}$ for large $lb_1$ to show the Hartman 
effect. Fig.~\ref{1barrier}(b) shows the phase time $\tau_2$ in arm $S_2$ 
which does not contain any barrier. This also evolves and saturates with 
$lb_1$, the length of the barrier in the other arm $S_1$. This delay 
is due to the 
contribution from paths which undergo multiple reflection in the first
branch before entering the second branch via junction point $J$. In absence 
of a barrier in the $n$-th arm the phase 
time $\tau_n$ measured close to the junction $J$ should go 
to zero {\em i.e.} $\tau_n \to 0$ in the absence of multiple scatterings in 
the first arm. 
Note that $\tau_{1s}$ and $\tau_{2s}$ change with energies of the incident
particle (Fig.~\ref{1barrier}). From Fig.\ref{1barrier} it can be easily
seen that $\tau_{2s}$ is always smaller than $\tau_{1s}$ for any particular 
$V_1$ {\em i.e.} the saturation time in the arm having no barrier is smaller. 
The phase time in both the arms show non-monotonic behavior as a function of
$V_1$. As we decrease the strength of the barrier $V_1$ the value 
of $\tau_1$ ($\tau_2$) decreases in the whole range of lengths of the barrier 
and also the saturated value of $\tau_{1s}$ ($\tau_{2s}$) decreases until 
$V_1$ reaches $1.6$ and with further decrease in $V_1$ the values of 
$\tau_1$ ($\tau_2$) as well as $\tau_{1s}$ ($\tau_{2s}$) starts increasing.

\begin{figure}[t]
\begin{center}
\includegraphics[width=10.0cm]{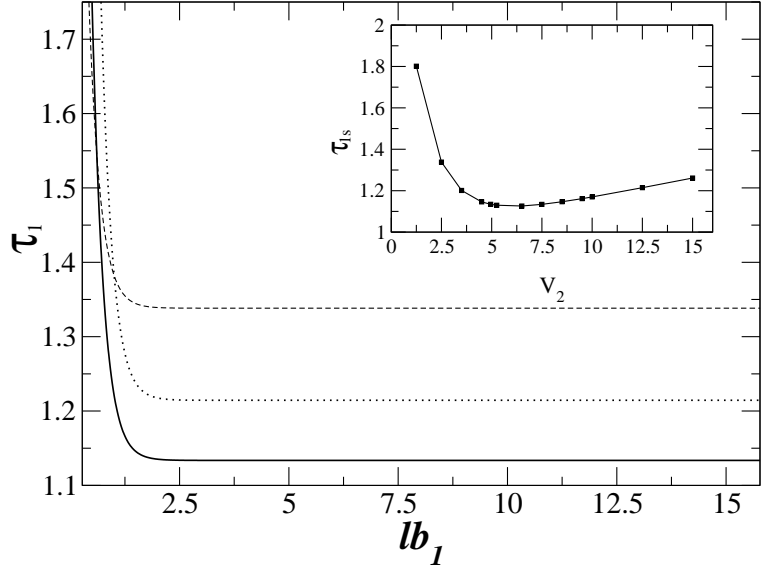}
\end{center}
\caption{Here for a 3-way splitter with one barrier in each branched 
arm $S_1$ and $S_2$, the `phase time' $\tau_1$ is plotted as a function 
of barrier length `$lb_1$' keeping $lb_2 (=1)$ and $V_1 (=5)$ fixed and for 
different values of parameter $V_2$. The 
dashed, solid and dotted curves are 
for $V_2 = 2.5, 5.0$ and $12.5$ 
respectively. Other system parameters are $E=1, w_1 = w_2 = 3$. In the
inset $\tau_{1s}$ is plotted as a function of $V_2$ for the same system
parameters.} 
\label{nonlocal-hartman}
\end{figure}

As the second case we take up another Y-junction which contain potential 
barriers in both of its side branches as shown in Fig.~\ref{system}. We fix the
values of $V_1 (=5)$ and vary $lb_1$ for each values of $V_2$
to study the $lb_1$-dependence of $\tau_1$ (Fig.\ref{nonlocal-hartman}). 
From Fig.~\ref{nonlocal-hartman} we see that $\tau_1$ decreases 
with increase in $lb_1$ and saturate to a value $\tau_{1s}$ at each value of 
$V_2$ thereby showing `Hartman effect' for arm `$S_1$'.
But now, we can tune the saturation phase time i.e. speed of the peak of the 
wave packet in one arm $S_1$ non-locally 
by tuning the strength of the barrier potential $V_2$ sitting on the other arm 
$S_2$! {\it Thus `quantum nonlocality' enables us to control the `super arrival' time
in one of the arms ($S_1$) by changing a parameter ($V_2$) on the other, 
spatially separated from it}. In the inset of Fig.~\ref{nonlocal-hartman} we
plot $\tau_{1s}$ as a function of $V_2$. It clearly shows that when 
the barrier strengths $V_1$ and $V_2$ are very close the `phase time' 
reaches 
its minimum value. In all other cases {\em i.e.,} whenever $V_1 \ne V_2$, the 
value of $\tau_{1s}$ is larger.

\begin{figure}[t]
\begin{center}
\includegraphics[width=10.0cm]{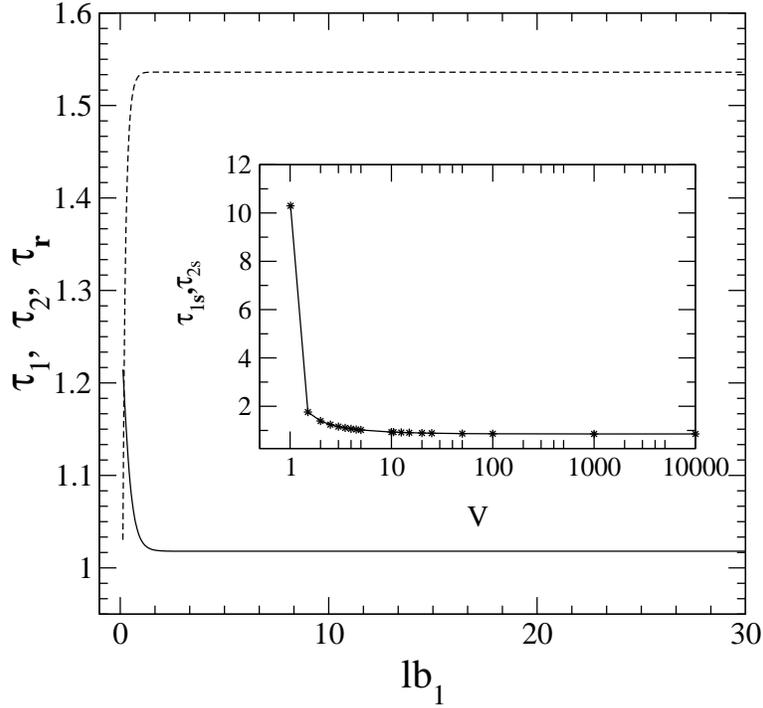}
\end{center}
\caption{Here for a 3-way splitter with an identical barrier in each branched 
arm $S_1$ and $S_2$, the `phase times'  
are plotted as a function of barrier length `$lb_1$'$(=lb_2)$ for fixed 
$V_1=V_2 (=5)$. The solid and dashed curves represent 
$\tau_1$ (=$\tau_2$) and $\tau_r$ 
respectively. Other system parameters are $E=1, w_1 = w_2 = 2.5$. In the
inset $\tau_{1s}(=\tau_{2s})$ is plotted as a function of $V(=V_1=V_2)$ 
for the same system parameters.} 
\label{nufig1}
\end{figure}

With the above system but for two identically branched arms (i.e., $V_1=V_2$, 
$w_1=w_2$ and $lb_1=lb_2$) we study the behaviour of different phase times as
a function of $lb_1$. For this case we see from Fig.~\ref{nufig1}  that 
$\tau_1$ and $\tau_2$ are the same, as expected. The phase time in 
reflection mode, measured near the junction on base arm,  
evolves differently from these two phase times. The reflection phase time 
getting saturated with increasing barrier's length confirms Hartman effect. 
The value of saturated reflection phase time is also different from that of 
the transmission mode. We also study the 
saturated transmission phase times by changing the strength of the barriers.
From the inset of Fig.~\ref{nufig1}, we see that for strength very close 
to the value of the incident energy the saturated phase time is quite large and 
the saturated phase time decreases with the increasing barrier strength.

\begin{figure}[t]
\begin{center}
\includegraphics[width=10.0cm]{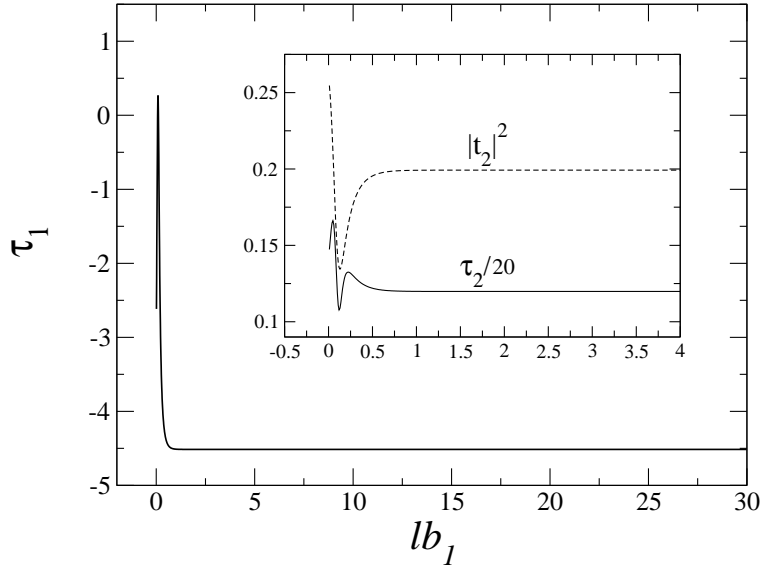}
\end{center}
\caption{Here for a 3-way splitter with one barrier in each side branch $S_1$
 and $S_2$, the `phase time' $\tau_1$ is plotted as a function 
of `$lb_1$' for a very small 
$lb_2 (=0.5)$. Other system parameters are $E=1$, $w_1=w_2=2.5$, $V_2=5$ 
and $V_1=15$. In the inset, the solid and dashed 
curves represent $\tau_2$ and $|t_2|^2$ respectively as a function of $lb_1$. For better visibility we have plotted phase times scaled down by a factor of $20$.} \label{2bar-negative}
\end{figure}

We shall now show  another interesting result related to the Hartman effect. 
For this we keep $V_2 (=5)$ unaltered and reduce $lb_2$.  
For very small $lb_2 (=0.5)$ we see from Fig.~\ref{2bar-negative} that 
$\tau_1$ is negative for almost the whole range of $lb_1$-values
 showing `time-advancement' and eventually after a sharp decrease saturates 
to a negative value of $\tau_{1s}=-4.514$  
implying `Hartman effect' with advanced time or `ultra Hartman effect'.
 In the inset we plot the corresponding $\tau_2$ and $|t_2|^2$ as a function 
of $lb_1$. 
The tunneling phase time for an one dimensional barrier with the same strength 
$V_2$ and width $lb_2$, as used in the side branch $S_2$ of the network, is
different from $\tau_2$, obtained for the whole range of values of $lb_1$. 
This is due to the presence of the other arm in the network problem. In the
cases discussed so far $\tau_2$ vary more sharply in small $lb_1$
regime. Furthermore, the inset in Fig.~\ref{2bar-negative} shows a dip in 
$\tau_2$ at parameter regimes where $|t_2|^2$ has a minimum. 

\begin{figure}[t]
\begin{center}
\includegraphics[width=10cm]{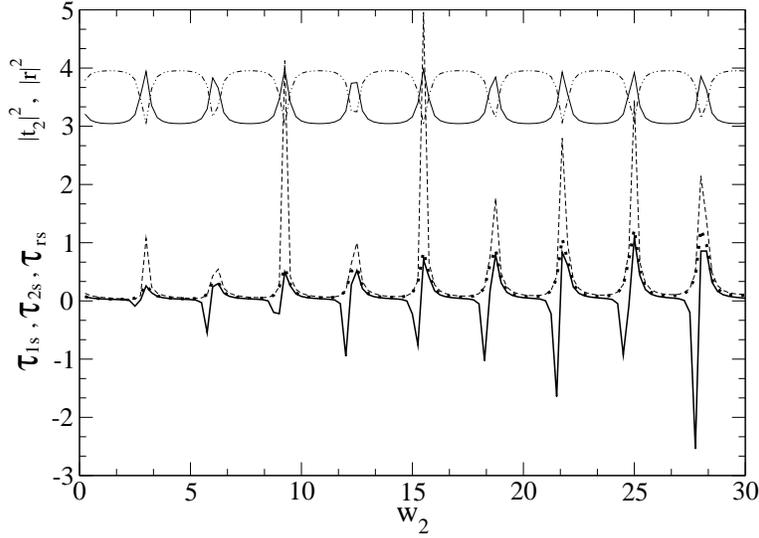}
\end{center}
\caption{For a 3-way splitter with one barrier in each side branch $S_1$
 and $S_2$, $\tau_{1s}/50$ (thick solid), $\tau_{2s}/50$
(thick dotted) and $\tau_{rs}/50$ (dashed) are shown as a function
of `$w_2$'. 
$|t_2|^2$ (solid) and  $|r|^2$ 
(double dot-dashed) are shifted upwards along the the positive $y$-direction 
by 3. The different system parameters are $E=1$, $V_1=15$, $V_2=5$, 
$lb_1=100.0$, $lb_2=0.5$ and $w_1=2.5$.} 
\label{negtaus}
\end{figure}

For a wave packet 
with large spread in real space it is 
possible that the leading edge of the wave packet reaches the barrier much
earlier than the peak of the packet. This leading edge in turn can tunnel 
through the barrier to produce a peak in the other end much before the 
peak of the incident wave packet reaches the barrier region. This 
sometimes is referred to as pulse 
reshaping effect. This, in general, causes `time advancement'\cite{landauer94}.
This negative delay does not violate causality, however, this delay time 
is bounded 
from  below. In 1D barrier such a situation does not arise. In  
presence of square wells in one dimensional systems 
negative time delays have been observed. This effect is termed as 
`ultra Hartman effect' [ see for details\cite{muga02} ]. 

\begin{figure}[t]
\begin{center}
\includegraphics[width=10cm]{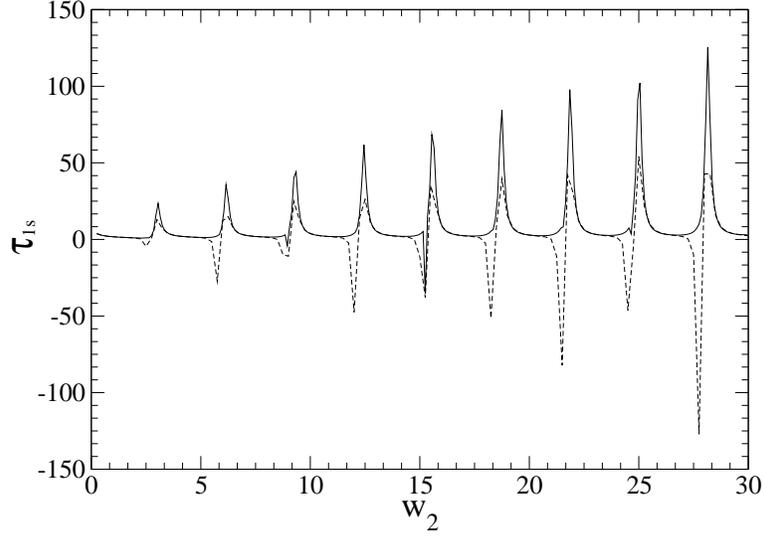}
\end{center}
\caption{Here for a 3-way splitter with one barrier in each side branch $S_1$
 and $S_2$, the `saturated phase time' $\tau_{1s}$ is plotted as a function of 
`$w_2$'. The dashed and solid curves are for $lb_2=0.5$ and $2.0$ respectively.
 Other system parameters are $E=1$, $V_1=15$, $V_2=5$, $lb_1=100.0$ and 
$w_1=2.5$.} 
\label{negtau1s}
\end{figure}

As the next case we set the length of the barrier in arm
$S1$ of the above mentioned system at a large value (say 100) 
where all the phase times get saturated.
Now we shift the position of the barrier in arm $S2$ away from the junction and
study its effect on the saturated transmission and reflection phase times. In
Fig.~\ref{negtaus} we have plotted all these three quantities $\tau_{1s}$, 
$\tau_{2s}$ and $\tau_{rs}$ as a function $w_2$. Earlier we had shown in
Fig.~\ref{nonlocal-hartman} that by changing a nonlocal parameter $V_2$ one can
tune $\tau_{1s}$ whereas Fig.~\ref{negtaus} shows that change of another 
parameter
$w_2$ can tune $\tau_{1s}$ nonlocally. Note that $\tau_{2s}$ and $\tau_{rs}$
also depend on $w_2$. From Fig.~\ref{negtaus} we see that $|t_2|^2$ ($|r|^2$)
shows resonances (anti-resonances) as a function of $w_2$. 
The saturated delay time $\tau_{1s}$ exhibit sharp variation in phase  
time around the resonance. Moreover, it takes negative as well as positive 
values. Around the resonance sharp peaks exist in both positive and negative 
side. In contrast to this observation of $\tau_{1s}$ the other two phase times
show only positive peaks. As $|t_2|^2$ and $|r|^2$ have finite
non-zero values, these variations in phase times should, in principle, be 
observable in experiments. In Fig.~\ref{negtau1s} we have plotted $\tau_{1s}$ 
as a function of $w_2$ for two different values of the length $lb_2$ of the 
barrier in arm $S2$. Note that as we increase the length $lb_2$, the frequency 
of getting negative saturation values ($\tau_{1s}$) reduces (see the solid 
curve in Fig.~\ref{negtau1s}) and increasing the length $lb_2$ further, the 
negative saturation goes away. This is in agreement with the discussions in 
previous paragraph. The saturation to a negative value can be attributed to
the nonidentical barriers in two branched arms, the barrier length of one 
being very small compared to the other. When $lb_1=lb_2$ the time-advancement goes away.

So far our discussion is based on wave guide theory of transport.
We can as well treat the junction as a scatterer described by the 
following one parameter scattering matrix\cite{buttiker85}
\be
S_J = \left(\begin{array}{ccc}
-(a + b ) & \sqrt{\epsilon} & \sqrt{\epsilon}\\
\sqrt{\epsilon} & a & b\\
\sqrt{\epsilon} & b & a
\end{array}\right)
\label{S-J}
\ee
where $a = \frac{1}{2} \left( \sqrt{1-2\epsilon} - 1 \right)$ and 
$b = \frac{1}{2} \left( \sqrt{1-2\epsilon} + 1 \right)$. $\epsilon$ is
a coupling parameter with values $ 0 \le \epsilon \le 0.5$. When
$\epsilon \to 0$ the side branches are decoupled from the base arm 
while for $\epsilon \to 0.5$ they are strongly coupled. This $S$-matrix
satisfies the conservation of current\cite{shapiro83,datta}. For our network
system with two side branches, the equations at the junction $J$ can be 
written as
\be
\left(\begin{array}{c}R \\ A_1 \\ A_2 \end{array}\right) = S_J\,
\left(\begin{array}{c}1 \\ B_1 \\ B_2
\end{array}\right)\,.
\label{eq:S-J}
\ee

\begin{figure}[t]
\begin{center}
\includegraphics[width=10cm]{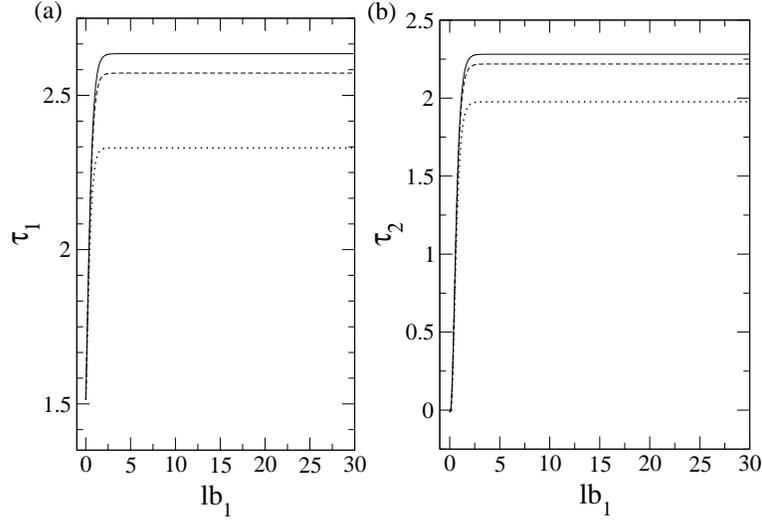}
\end{center}
\caption{For a 3-way splitter with a barrier in $S_1$ arm, the `phase times' $\tau_1$ and $\tau_2$ are plotted as a function of barrier length `$lb_1$' in $(a)$ and $(b)$ respectively. The dotted, dashed and solid curves are for the junction coupling parameter $\epsilon =1/9, 1/3, 4/9$ respectively. Other system parameters are $ V_1=3, E = 1, w_1 = 3$. We see that both $\tau_{1s}$ and $\tau_{2s}$
increase as we increase the coupling strength.} 
\label{cpling}
\end{figure}

We have studied the effect of the coupling parameter $\epsilon$ on the 
Hartman effect. From
Fig.~\ref{cpling}, we see that though the nature of $\tau_1$ and 
$\tau_2$ as a function of $lb_1$ does not depend on the coupling parameter 
$\epsilon$ but their saturation value is strongly dependent on $\epsilon$. 
 For weak coupling, the value of saturated phase
time is quite low and it increases with increasing strength of the coupling 
between base arm and side branches. For $\epsilon=4/9$ we get back the results 
obtained earlier using the wave-guide theory (compare the solid curves in 
Fig.~\ref{cpling} with the dashed curves in Fig.~\ref{1barrier}).

\section{Hartman effect in Quantum ring}
\label{time2}
\begin{figure}[t]
\begin{center}
\includegraphics[width=10.0cm]{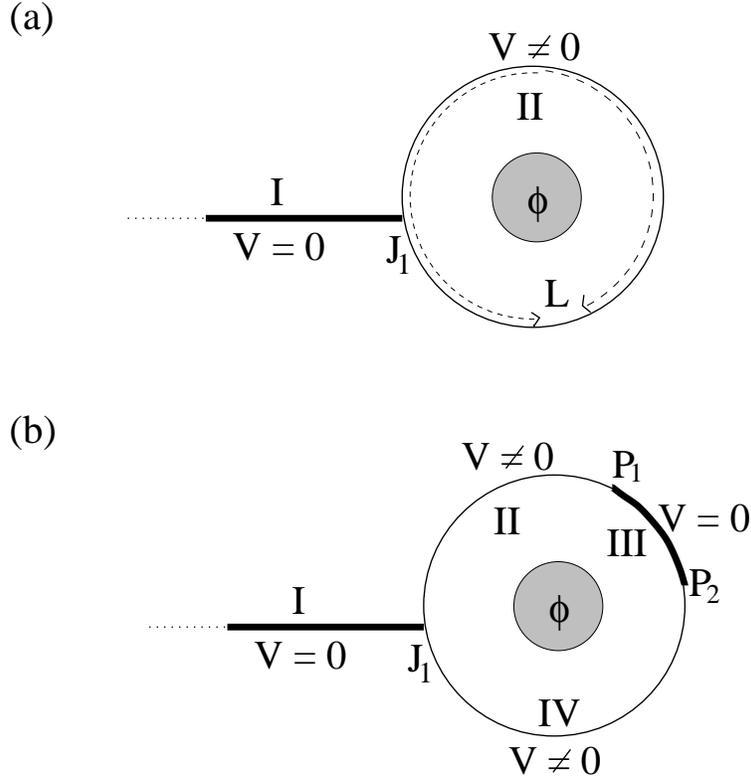}
\caption{Schematic diagram of a circular ring connected to semi infinite lead.} 
\label{1leadsys}
\end{center}
\end{figure}

In this section we study the scattering problem across a quantum ring 
connected to one ideal semi infinite lead (as shown schematically in 
Fig.~\ref{1leadsys}). 
Such ring geometry systems are extensively investigated 
in mesoscopic physics in analyzing normal state Aharanov-Bohm 
effect which has been observed experimentally\cite{washburn86,gefen84}. 
A magnetic field is applied perpendicular to the plane of the ring.
Due to this a magnetic flux $\phi$ as shown in Fig.~\ref{1leadsys},  
enclosed by the ring, there is a finite quantum mechanical potential of 
strength $V$ inside the ring while that in the connecting 
lead is set to be zero. We focus on a situation wherein 
the incident electrons have an energy $E$ less than $V$. 
The impinging electrons in this sub-barrier 
regime travels as an evanescent mode throughout 
the circumference of the ring and the reflection or the conductance 
involve contributions from both the Aharanov-Bohm effect as well as 
quantum tunneling. Here we are interested in a single channel case where the 
Fermi energy lies in the lowest sub-band. To excite the evanescent modes 
in the ring we have to make the 
width of the ring much less than that of the connecting lead. The electrons 
occupying the lowest sub-band in the lead on entering the ring 
experience a higher barrier (due to higher quantum zero point energy)
and propagate in the ring as evanescent mode. The transmission or conductance 
across such systems has been studied in detail\cite{jayan94mpl,gupta}. An 
analysis of the phase time for such a ring system is carried out
in the following subsections.

Now we discuss the `reflection phase time' for the above
quantum ring in presence of AB-flux as shown in 
Fig~\ref{1leadsys}(a). It is well known that in one dimensional
  scattering/tunneling problem, reflection involves prompt part as well
  as the multiple scattering arising from the edges of the
  scattering center (say, for the square barrier).
  However, transmission across the scattering region does not have
  the prompt part but has only contributions from multiple scattering.
  We would like to emphasize here the fact that the unitarity of the
  scattering matrix forces transmission and reflection phase
  times to coincide for a one dimensional tunneling problem (to be identical 
in magnitude),
  even though reflection has a prompt part as mentioned above. 
Hence, 
the information we get does not depend on whether we study the phase time 
in the reflection or in the transmission mode. Thus in the present section 
we have chosen a simple and generalized geometry where we can analytically 
study the phase time in a reflection mode and in presence of 
Aharonov-Bohm flux.  
We show that this phase time in the opaque barrier 
regime becomes independent of the length of the circumference of the ring
and the magnitude of the AB-flux. We have also studied this effect by
including an additional potential well between two barriers in the circular 
ring (Fig.~\ref{1leadsys} (b)). Interestingly, 
the saturated reflection phase time becomes independent of the length of the 
potential well (in the large length limit) for energy away from resonances. 
Inside the potential well the electron travels with a real velocity. 
Increasing or decreasing the free path (length of the well) does not alter 
the saturated reflection phase time through the system. It seems as if 
for the electronic wave the free space collapses. This result is
regarded as a ``space collapse" or ``space destroyer"\cite{recami02}. 

\subsection{Theoretical Treatment}
We use the same quantum wave guide theory\cite{xia92,jayan94mpl} as 
discussed in the earlier section. To set the equations in most general 
context we consider Fig.~\ref{1leadsys}(b) in presence of flux $\phi$.
The wave functions in different regions are
\begin{eqnarray}
\psi_0(x_0) &=& e^{ikx_0} + r e^{-ikx_0}\,\,\,\,\,\mbox{( in region I )}\label{wv5}\\
\psi_1(x_1)&=& A_1\, e^{-\kappa_1\,x_1} + B_1\, e^{\kappa_1\,x_1}\,\,\,\,\, \mbox{( in region II )} \label{wv6}\\
\psi_w(x_w) &=& C\, e^{i\,k\,x_w} + D\, e^{-i\,k\,x_w} \,\,\,\,\, \mbox{( in region III) }\label{wv7}\\
\psi_2(x_2) &=& A_2\, e^{-\kappa_2\,x_2} + B_2\, e^{\kappa_2\,x_2}\,\,\,\,\, \mbox{( in region IV) }\label{wv8}
\end{eqnarray}
with $k$ being the wave-vector of electrons in the lead and in the 
intermediate free space between two barriers inside the ring. 
$\kappa_1=\sqrt{2m(V_1-E)/\hbar^2}$ and $\kappa_2=\sqrt{2m(V_2-E)/\hbar^2}$ 
are the imaginary wave-vectors respectively for tunneling electrons in the 
barriers of strength $V_1$ and $V_2$ inside the ring. The origin of the 
co-ordinates of $x_0$ and $x_1$ is assumed to be at $J_1$ and that for 
$x_w$ and $x_2$ at $P_1$ and $P_2$ respectively. At $P_1$, $x_1=lb_1$, at
$P_2$, $x_w=w$ and at $J_1$, $x_2=lb_2$, where $lb_1$ and $lb_2$ are the
length of the two barriers  and $w$ is the length of the well inside 
the ring. Total circumference of the ring is $L=lb_1 + lb_2 + w$.

In presence of the AB-flux, following the same method described above,
 the boundary conditions for the current system (shown in 
Fig.~\ref{1leadsys}(b)) are 
\begin{eqnarray}
1+r-A_1-B_1 \exp(-i\,\alpha_1)=0\, ,\label{eq:1leadbc1}\\
A_2\,\exp(-\kappa_2\,lb_2)\,\exp(i\alpha_2)+B_2\,\exp(\kappa_2\,lb_2)\nn\\
-1-r=0\, ,\\
ik\,(1-r)+\kappa_1\,(A_1-B_1\,\exp(-i\,\alpha_1))-\,\kappa_2\,A_2\,\nn\\
\exp(-\kappa_2\,lb_2)\,\exp(i\,\alpha_2)
-\,\kappa_2\,B_2\,\exp(\kappa_2\,lb_2)=0\, ,\\
A_1\,\exp(-\kappa_1\,lb_1)\,\exp[i\,\alpha_1]+B_1\,\exp(\kappa_1\,lb_1)\nn\\
-\,C-D\,\exp(-i\,\alpha_w)=0\, ,\\
\kappa_1\,A_1\,\exp(-\kappa_1\,lb_1)\,\exp(i\,\alpha_1)-\kappa_1\,B_1\,\exp(\kappa_1\,lb_1)\nn\\
+\,i\,k\,C-i\,k\,D\,\exp(-i\alpha_w)=0\, ,\\
C\,\exp(i\,k\,w)\,\exp(i\,\alpha_w)\,+\,D\,\exp(-i\,k\,w)\nn\\
-\,A_2-B_2\,\exp(-i\alpha_2)=0\, ,\\
i\,k\,C\,\exp(i\,k\,w)-i\,k\,D\,\exp(-i\,k\,w)\,\exp(-i\alpha_w)\,\nn\\
+\,\kappa_2\,A_2-\kappa_2\,B_2\,\exp(-i\,\alpha_2)=0\, ,\label{eq:1leadbc7}
\end{eqnarray}
where $i\,\alpha_1$, $i\,(\alpha_1+\alpha_w)$ are the phases picked up 
respectively at $P_1$ and $P_2$ by the electron traveling clockwise 
from $J_1$ and $i\,(\alpha_1+\alpha_w+\alpha_2)$ is the phase picked up
by the same electron at $J_1$ after traversing once along the ring. 
The total phase around the ring becomes 
$\alpha_1\,+\,\alpha_w\,+\,\alpha_2\,=\,2\,\pi\,\phi/\phi_0$. 

To obtain an analytical expression for the reflection amplitude for a ring 
system as shown in Fig.~\ref{1leadsys}(a) we solve 
Eqs.~(\ref{eq:1leadbc1})-~(\ref{eq:1leadbc7}) 
using $C=0$, $D=0$, $A_2=A_1$, $B_2=B_1$, $\kappa_2=\kappa_1$ in the wave 
functions~(\ref{wv7}) and ~(\ref{wv8}) and $x_1=L$ at $J_1$ after a complete
traversal along the circumference. The reflection amplitude is
\begin{equation}
r=\frac{-\kappa_1\,(2\,\cos(\alpha)-\exp(kL))+i\,\frac{k}{2}\,\exp(kL)}
{\kappa_1\,(2\,\cos(\alpha)-\exp(kL))+i\,\frac{k}{2}\,\exp(kL)}\, ,\label{ref}
\end{equation} 
where $\alpha=2\,\pi\phi/\phi_0$. After knowing $r$, the
 `reflection phase time' $\tau_r$ can be calculated from 
the energy derivative of its phase\cite{hauge89,wigner55} as
\begin{equation}
\tau_r\, = \, \frac{\partial Arg[r]}{\partial E}\, . \label{refph}
\end{equation}
\subsection{Results and Discussions}

\begin{figure}[t]
\begin{center}
\includegraphics[width=10.0cm]{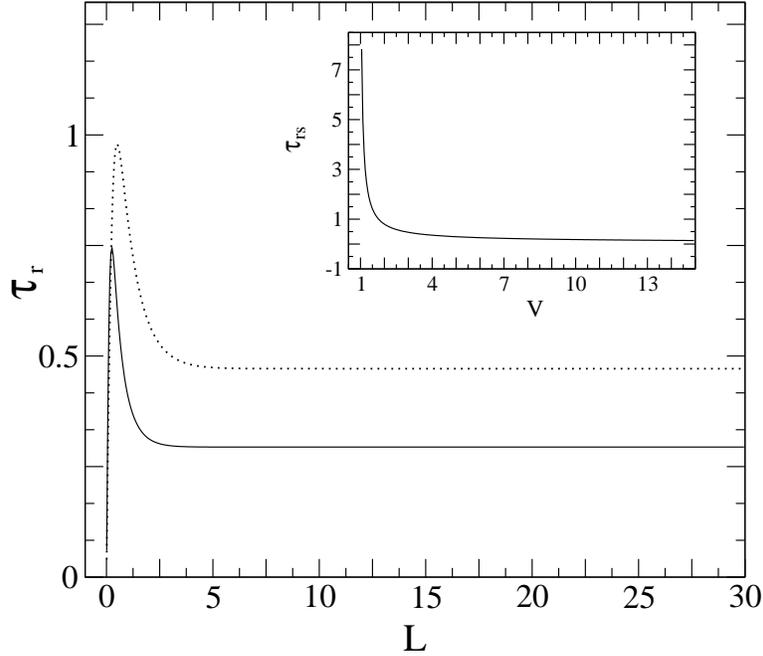}
\caption{In absence of magnetic flux ({\it i.e.} $\phi$ =0 ),  
for a ring with a barrier of strength $V$ throughout its circumference,
 the reflection phase time $\tau_r$ is plotted as a function of ring's 
circumference $L$.
 The solid, and dotted curves are for 
$V = 5, \, 3$ respectively. Incident energy is set to be 
$E=1$. In the inset the saturated value of phase time $\tau_{r\,s}$ 
is plotted as a function of the barrier's strength for same $E$. 
} 
\label{fig:tau-L}
\end{center}
\end{figure}

We now proceed to analyze
the behavior of $\tau_r$ as a function of various 
physical parameters for these ring systems. In the similar fashion, 
described above, here also we express all the physical quantities in 
dimensionless units. Thus the reflection phase time $\tau_r$ is expressed
in units of inverse of incident energy $E$ ($\tau_r\,\equiv\,E\,\tau_r$). 
After straight forward algebra in the large length ($L$) limit and in absence 
of magnetic flux, we obtain an analytical expression for the saturated 
reflection phase time (using Eq.~(\ref{ref}) in Eq.~(\ref{refph})), which is 
given by,
\begin{equation}
\tau_{r\,s}=\frac{\frac{1}{k\,\kappa}+\frac{k}{\kappa^3}}{\left(2\,+\,\frac{k^2}{2\kappa^2}\right)}\, ,\label{satphtm}
\end{equation} 
with $\kappa$ being the imaginary wave vector of the electron inside the 
barrier of strength $V$ and length $lb\,\left(=L\right)$.

First we take up a ring system with a single barrier along the circumference 
of the ring as shown in Fig.~\ref{1leadsys}(a). For a tunneling particle 
having energy less than the barrier
strength we find out the reflection phase time $\tau_r$ as a function of 
barrier length $L$ which in turn is the circumference of the ring. We see 
(Fig.~\ref{fig:tau-L}) that in absence of magnetic flux, $\tau_r$ evolves as 
a function of $L$ and asymptotically saturates to a value $\tau_{r\,s}$ which 
is independent of $L$ thus confirming the `Hartman effect'. From 
Fig.~\ref{fig:tau-L} it is clear that the saturation value increases with the 
decreasing barrier-strength. In the inset of Fig.~\ref{fig:tau-L}, we have 
plotted $\tau_{r\,s}$ as a function of barrier-strength. From this we can 
see that for electrons with incident energy close to the 
barrier-strength the value of $\tau_{r\,s}$ is quite large. 

\begin{figure}[t]
\begin{center}
\vskip 1cm
\includegraphics[width=10cm]{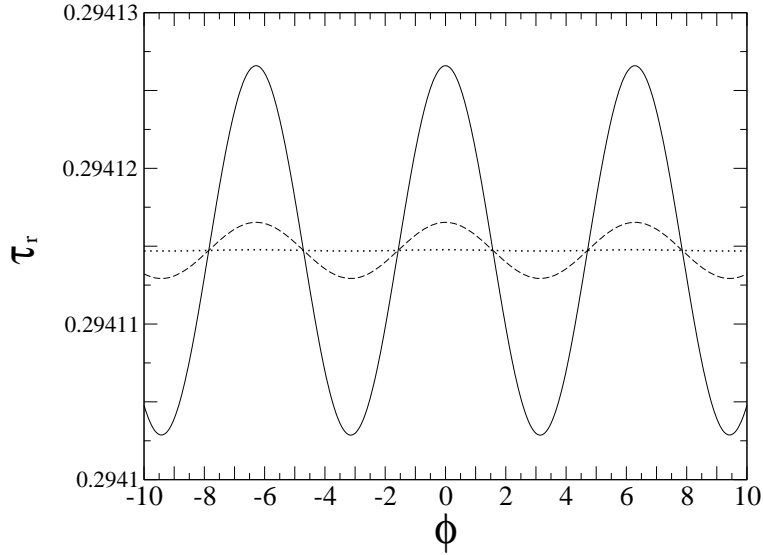}
\end{center}
\caption{For a ring with a barrier of strength 
$V$ lies throughout its circumference, the saturated phase time 
$\tau_{r\,s}$ is 
plotted as a function of magnetic flux $\phi$. The solid, dashed and dotted 
curves are for $L = 6, 7, 9$ respectively. Other system parameters are $V=5$,
$E=1$.} 
\label{fig:tau-phi}
\end{figure}

To see the effect of magnetic flux on `Hartman effect', we consider the same
system but in presence of AB-flux. We study the reflection phase time as a function of AB-flux for different values of the length $L$ of the
ring's circumference. We have chosen the lengths such that in the  absence 
of the `AB-flux', for a given system ({\it i.e.,} for known $E$ and $V$) the 
reflection phase time $\tau_r$ gets saturated for these lengths. From 
Fig.~\ref{fig:tau-phi} we see that  $\tau_r$ as a function
of $\phi$ shows AB-oscillations with an average value which is the saturation
value $\tau_{r\,s}$ for the same system in absence of AB-flux. Further we 
observe that (Fig.~\ref{fig:tau-phi}) $\tau_r$ is flux periodic with 
periodicity 
$\phi_0$. This is consistent with the fact that all the physical properties in 
presence of AB-flux across the ring must be periodic function of the flux with 
a period $\phi_0$\cite{buttiker93,washburn86,datta}. However, we see that as 
we increase $L$ the magnitude of AB-oscillation in $\tau_r$ decreases. 
Consequently in the large length limit the visibility of oscillations vanishes. This clearly 
establishes `Hartman effect' even in presence of AB-flux. The constant value 
of $\tau_r$ thus obtained in the presence of flux is identical to 
$\tau_{r\,s}$ ($0.294115$) in the absence of flux 
(see Fig.~\ref{fig:tau-phi}) in the 
large length regime and its magnitude is given by Eq.~(\ref{satphtm}). This 
result clearly indicates that the reflection phase time in the presence of 
opaque barrier becomes not only independent of the length of the circumference 
but also is independent of the AB-flux thereby observing the `Hartman effect' 
even in the presence of AB-flux. Since the magnetic field tunes the boundary 
conditions and in the large length limit the evanescent states become 
insensitive to the boundaries (as they decay out before reaching 
the boundary) the Hartman effect follows. 

\begin{figure}[t]
\begin{center}
\includegraphics[width=10.0cm]{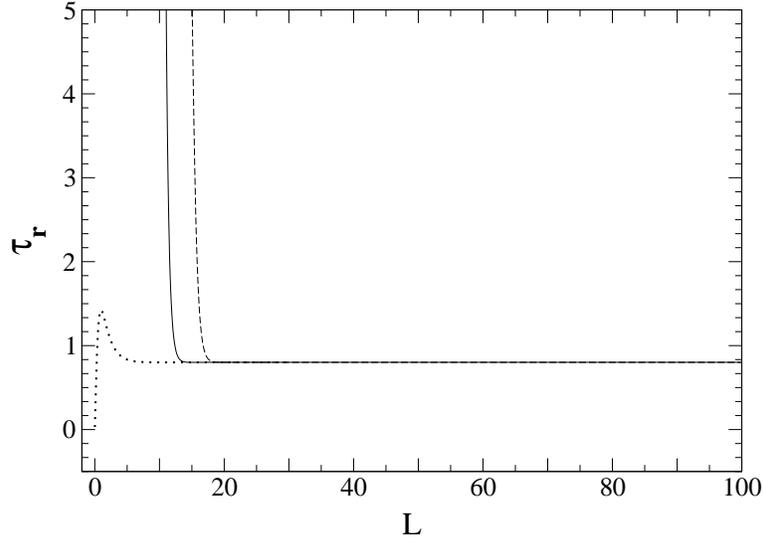}
\end{center}
\caption{In absence of magnetic flux ({\it i.e.} $\phi$ =0 ),  
for a ring with two barriers of strength $V_1$ and $V_2$ separated by an 
intermediate well region,
 the reflection phase time $\tau_r$ is plotted as a function of ring's 
circumference $L$ for different length $w$ of the well. The dotted, 
solid and dashed curves are for $w=0, 5, 10$ respectively. Other system 
parameters are $lb_2=5$, $V_1=V_2=2$ and $E=1$.} 
\label{fig:tau-w}
\end{figure}

Now we consider the ring system with the ring having two successive barriers 
separated by an intermediate free space (we shall call it `quantum well' in
what follows) as shown in Fig.~\ref{1leadsys}(b). In 
the absence of magnetic flux, we see the effect of `quantum well' on the 
reflection phase time $\tau_r$. In Fig.~\ref{fig:tau-w} $\tau_r$ is plotted as 
a function of one of the barrier length (say $lb_1$) while other barrier 
length is kept fixed ($lb_2=5$) and for few different values of length of the well. 
Here, the fixed value of $lb_2$ is chosen in such a way 
that in absence of the well region the reflection phase time reaches saturation at this length. From Fig.~\ref{fig:tau-w} we see that for all parameter values 
of well length, the saturation value of reflection phase time $\tau_{r\,s}$ 
is same and it is equal to what we obtained in absence of the well in the ring 
system. Thus the saturated phase time becomes independent of the length of the 
well (in the long length limit) for the energy away from the resonances. This 
is as if the 
effective velocity of the  electron in the well becomes infinite or 
equivalently length of the well
does not count (space collapse or space destroyer) while traversing the ring.

\begin{figure}[t]
\begin{center}
\includegraphics[width=10.0cm,angle=270]{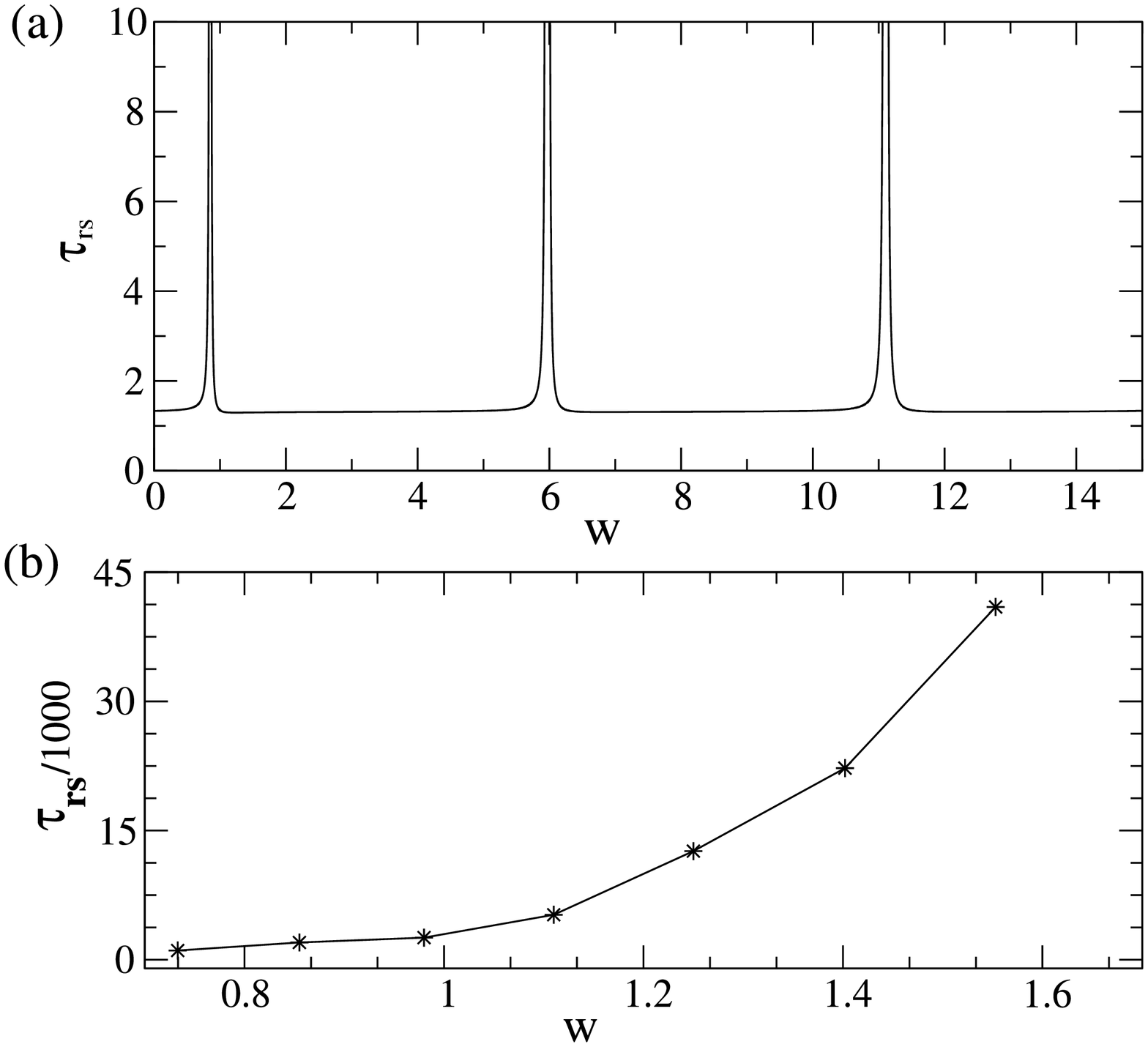}
\caption{$(a)$ In absence of magnetic flux ({\it i.e.} $\phi$ =0 ),  
for a ring with two identical barriers ($V_1 = V_2 =2$, $lb_1=lb_2=5$)
 separated by an intermediate well region of length $w$, the 
phase time $\tau_{r}$ is plotted as a function of $w$ 
at $E=1.5$. $(b)$ With change in incident energy the position of resonance peaks shift. We follow the first resonance peak and plot resonant phase time $\tau_r^*$ as a function of $w$ as we vary incident energy.
From left to right the points correspond to energies $E=1.6, 1.5, 1.4, 1.3, 1.2,1.1, 1.01$ respectively.}
 \label{fig:resonance}
\end{center}
\end{figure}

Next we consider a similar system as that shown in Fig.~\ref{1leadsys}(b).
Here we see the effect of resonances present in the ring system with the well, 
on the saturated 
reflection phase time $\tau_{r\,s}$. For the system described 
above with two identical barriers of strength $V_1=V_2=2$ and
length $lb_1=lb_2=5$ separated by an intermediate free space of
length $w$ we study the phase time $\tau_r$ as a function of $w$ in
absence of the magnetic flux ($\phi=0$). In Fig.~\ref{fig:resonance}$(a)$ 
we have plotted $\tau_{rs}$, for the 
electrons with an incident energy $E=1.5$. It should be noted that as we 
increase the length of the well the incident energy $E$ 
coincides with resonances (or resonant states) in the well (which arise due to 
constructive interference due to multiple scatterings inside the well). For 
these values of $w$ we observe sharp rise in the saturated delay time and 
its magnitude depends on the length of the well. We see that the 
resonances which have Lorentzian shape become broader as the 
length of the free region $w$ becomes large.  It is worthwhile to mention that 
away from the resonance the value of $\tau_{r\,s}$ is independent of the 
length of the well (see Fig.\ref{fig:tau-w}) and depends only on the barrier 
strength. The resonance being dependent on the incident energy of the
electrons we follow a particular resonance peak (say, the first peak) in
$\tau_{r\,s}$ for different $E$. In Fig.~\ref{fig:resonance}$(b)$, the first 
peak values are plotted as a function of $w$ for seven different incident 
energies in the sub-barrier tunneling regime. We see that as we increase the
incident energy the resonances shift towards lower $w$ value 
and also the peak-value decreases as expected. The above result clearly 
indicate that at and around the resonance the saturated value depends on the 
length of the free space, however, away from the resonance the saturated delay
time is independent of the length of the free space( space collapse or space 
destroyer as discussed earlier).
 
\begin{figure}[t]
\begin{center}
\includegraphics[width=10.0cm,angle=270]{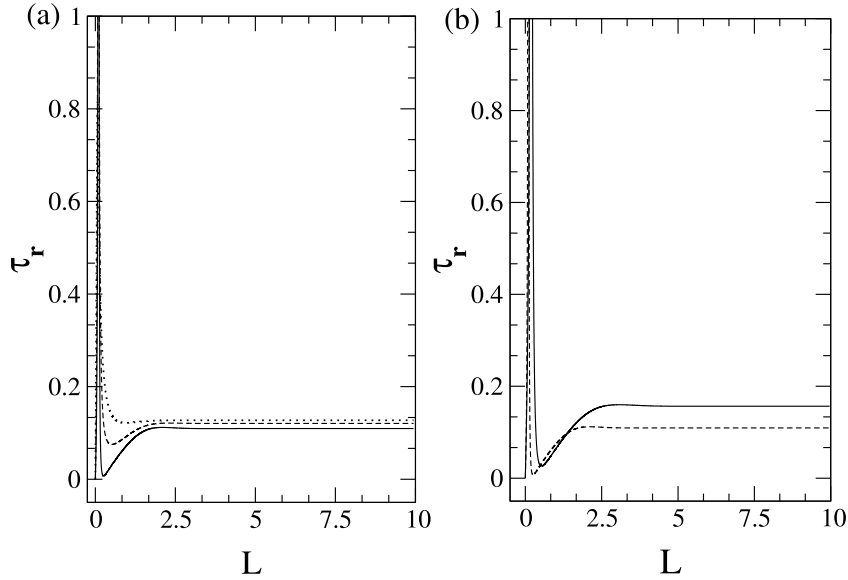}
\caption{In presence of a complex barrier covering the ring's circumference  
the phase time $\tau_r$ in reflection mode is plotted as a function of ring's
circumference. In $(a)$ the three different curves are for three different 
strength of the real part of the potential while the imaginary part is fixed 
at $V_{im}=10$. The solid, dashed and dotted curves are for $V_{re} = 2, 4$
and $10$ respectively. In $(b)$ the $\tau_r$ is plotted for two different 
values of the imaginary part of the potential keeping the strength fixed at
$V_{re} = 2$. The solid and dashed curves are for $V_{im}=5$ and $10$ respectively. Incident energy is set to be $E=1$.} 
\label{cmplxring}
\end{center}
\end{figure}

Instead of a real barrier  we now consider a complex potential (or absorption) 
$V=V_{re}\,-i\,V_{im}$in Fig.~\ref{1leadsys}$(a)$. Here, $V_{re}$ 
represents the real strength of the barrier whereas $V_{im}$ is the 
imaginary strength. In presence of the imaginary part in the potential 
our ring system can absorb particles in it. 
Earlier Raciti {\em et. al.}, have shown\cite{raciti} that the phase time in 
transmission mode for an absorptive one-dimensional rectangular barrier 
does not show the saturation in opaque barrier limit, instead, it is linearly
proportional to the length of the barrier. Thus they  conclude that
in presence of strong absorption the tunneling time does not show the 
Hartman effect. We have separately studied the phase time in both transmission 
and reflection modes of the one-dimensional complex barrier. We have found that 
in presence of non-negligible absorption in the large length
limit of the barrier though the transmission phase time does not saturate 
the reflection phase time shows Hartman effect. It would be worth 
to mention here that the saturated reflection phase time of 1D complex barrier 
is same as that for
an one-dimensional complex step potential. Now we study the phase time
in reflection mode (Fig.~\ref{1leadsys}$(a)$) in the presence of a complex
potential. We vary the barrier length (ring circumference) for
different values of $V_{re}$ and $V_{im}$. From Fig.~\ref{cmplxring}, we
see that even in presence of complex potential Hartman effect survives for
our ring system. In Fig.~\ref{cmplxring}$(a)$, keeping the imaginary part of 
the potential fixed we vary the strength of the barrier.  We  see that 
the saturated phase time as well as the nature of evolution
of the $\tau_r$ as a function of $L$ depend on $V_{re}$. From
Fig.~\ref{cmplxring}$(b)$, we see that the saturated reflection phase time
changes with $V_{im}$ as well.

\begin{figure}[t]
\begin{center}
\includegraphics[width=10.0cm]{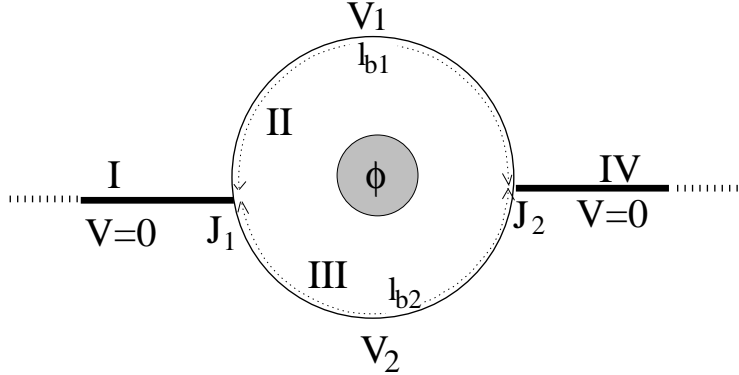}
\end{center}
\caption{Schematic diagram of a ring connected to two 
leads at the junctions $J_1$ and $J_2$ in the presence of an 
Aharanov-Bohm flux, $\phi$. The upper arm of the ring is covered 
by a barrier of strength $V_1$ and length $lb_1$. Similarly, the lower 
arm is covered by a barrier of strength $V_2$ and length $lb_2$. The 
circumference of the ring is $L=lb_1 + lb_2$.} 
\label{2leadsys}
\end{figure}

\begin{figure}[t]
\begin{center}
\includegraphics[width=10.0cm]{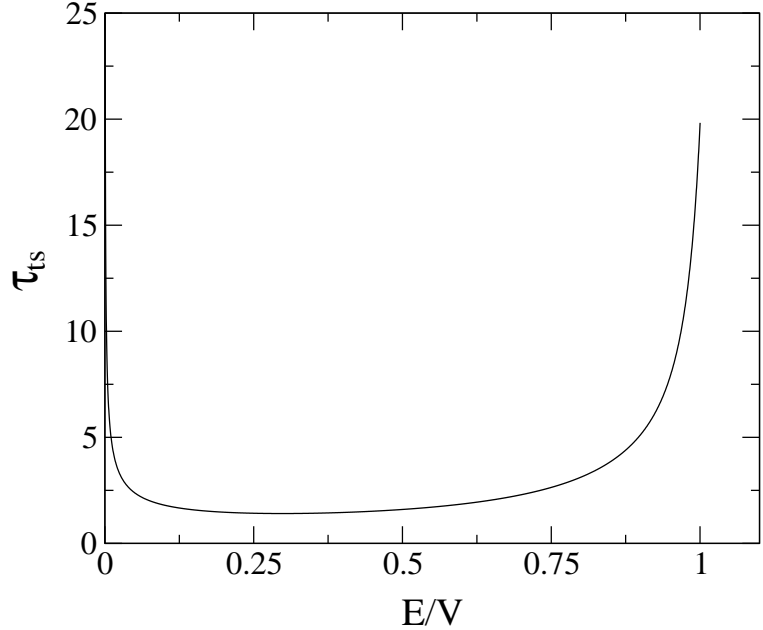}
\end{center}
\caption{In absence of magnetic flux ({\it i.e.} $\phi$=0),  
for a ring shown in Fig.~\ref{2leadsys} the saturated transmission 
phase time $\tau_{t\,s}$ is plotted as a function of incident energy $E/V$. 
The length of the identical barriers ($V_1 = V_2$, $lb_1=lb_2$) in the 
upper and lower arms of the ring are $L/2=15$.} 
\label{ring2lead}
\end{figure}

Finally we consider a quantum ring connected with two identical leads as shown 
in Fig.~\ref{2leadsys}. Each of the two arms (say upper and lower arm) of the ring are covered entirely by a 
barrier. We study the transmission phase time in the absence of magnetic 
flux by varying different system parameters. Here also the phase time 
evolves with 
increasing length of
the barriers and for larger length the phase time gets saturated. The 
saturation
value does not depend on the ratio of the length of the two arms of the ring. 
However, this value depends on the incident energy of the particles. 
In Fig.~\ref{ring2lead} we have plotted the saturated transmission 
phase time $\tau_{t\,s}$ as a function of incident energy $E/V$ for 
$\phi=0$. The circumference of the ring is taken as $L=30$ with equal upper 
and lower arm lengths.
Plots with different arm length ratios ($lb_1:lb_2$) with different $\phi$ 
in the asymptotic limit were
found to overlap with the above curve in the entire energy regime. 
Analytically, in the large $L$ ($>1/\kappa$) limit, the transmission phase time 
$\tau_t$ becomes independent of $L$ and the magnetic flux (in accordance with 
Hartman effect) and is given by
\begin{equation}
\tau_{t\,s}\,=\,
\frac{4\,\kappa^3\,+\,5\,k^2\,\kappa\,+\,(k^4/\kappa)}{2\,k\,((2\,\kappa^2\,-\,(k^2/2))^2+
4\, k^2 \,\kappa^2)}. \label{taus} 
\end{equation}

\section{Conclusions}\label{conclude}
We have verified the Hartman effect in quantum network 
consisting of an one dimensional arm having two side branches. 
These side branches may or may not have barriers. In presence of a barrier,
the `phase time' 
for transmission through a side branch shows the `Hartman effect'. 
Due to quantum nonlocality the `phase time' and its 
saturated value at any side branch feel the presence of barriers in other 
branches. Thus one can tune the saturation value of
`phase time' and consequently the superluminal speed in one branch 
by changing barrier parameters in any other 
branch which is spatially separated from the former. 
Moreover, Hartman effect with negative saturation times (time advancement) 
has been observed for some parameter values.  The phase times are also sensitive to the junction $S$-matrix elements used for a given problem.  
Depending on $w_n$ there may be one or several bound states located 
between the barriers in different branched arms and as a consequence
saturated delay time can be varied from the negative (ultra Hartman
effect) to positive and vice-versa. We have verified this by looking at the
transmission coefficient in the second arm $S_2$ which exhibits clear 
resonances as a function of $w_2$.  
Moreover the reported effects are 
amenable to experimental verifications in the electromagnetic wave-guide 
networks.

We have further extended our studies on Hartman effect in quantum ring 
geometry in the presence of AB flux. We have studied the quantum 
reflection phase time for a ring connected with a single lead. Our study show 
that the phase time for a given incident energy becomes independent of the 
barrier thickness as well as the magnitude of the flux in the limit of opaque 
barrier. In the absence of AB-flux 
we have obtained analytical expressions for the saturated reflection and 
transmission phase times for the ring geometries.  
In addition, introducing a potential well between two successive opaque 
barriers covering the entire circumference of the ring, we have found that the 
saturated reflection phase time becomes independent of the length of the well 
for energies away from the resonances. This implies, as if, the effective 
velocity of the electron within the well becomes infinite or equivalently the 
length of the well does not count (space collapse or space destroyer). 
Earlier studies reveal that the Hartman effect disappears in presence of complex barrier in a transmission mode. We have shown that the Hartman effect 
survives in the reflection mode.   

\section{Acknowledgments}
One of the authors (SB) thanks Doron Cohen and Debasish Chaudhuri 
for useful discussions.
SB also thanks SNBNCBS, Kolkata and IOP, Bhubaneswar where a 
large part of the present work was carried out.

 


\end{document}